# WSe$_2$ p-MOSFETs with Nb-Doped WS$_2$ Contacts Deposited using Atomic Layer Deposition

Shivanshu Mishra, Ruixue Li, Dongjea Seo, Anil Adhikari, Lucas G. Cooper, Rebecca A. Dawley, Ageeth A. Bol, and Steven J. Koester, *Fellow, IEEE*

*Abstract*—**WSe$_2$ p-MOSFETs with Nb-doped WS$_2$ contacts formed using atomic layer deposition are demonstrated. The devices are fabricated using a technique that aligns the contact metallization with the Nb-doped WS$_2$ contacts using a selective oxidation process. Devices with source/drain spacing of 0.15 μm have on-state current of 103 μA/μm at $V_{DS}$ = -1 V at a channel carrier concentration of ~ 7.5 × 10$^{12}$ cm$^{-2}$. The results provide a promising CMOS-compatible pathway to create low-resistance contacts to 2D-channel transistors.**

## I. INTRODUCTION

Two-dimensional (2D) transition metal dichalcogenides (TMDs) are promising semiconductors for future scaled CMOS technologies due to their atomically thin channels, which allow excellent electrostatic control and improved scalability compared to silicon [1]. Among these, tungsten diselenide (WSe$_2$) is particularly suited for p-channel devices since it offers a suitable bandgap (~1.2 eV in monolayer form), high hole mobility, and a relatively shallow valence band edge, making p-type contacts more achievable [2].

Despite significant progress on WSe$_2$ p-MOSFETs, device performance remains limited by high contact resistance. Reliable hole injection is difficult due to strong Fermi-level pinning from metal-induced gap states and the lack of suitable semi-metallic contact materials. Techniques such as surface doping [3], insertion of interfacial layers [4], and use of high–work function metals [5] or oxides [6] have improved performance but often suffer from poor scalability, limited thermal stability, or incompatibility with CMOS processing [7].

A promising route to low-resistance contacts is Nb-doped layers [8]. Substitutional Nb doping during channel growth can produce degenerate p-type behavior with hole densities up to 10$^{19}$–10$^{20}$ cm$^{-3}$ [9]. However, direct channel doping does not allow independent control of the contact resistance, $R_C$, and threshold voltage. Another strategy is 2D/2D contact stacks using metallic TMDs such as NbSe$_2$, TaS$_2$, or alloyed WSe$_2$ [10],[11]. These methods highlight the benefits of heavy p-type doping at the contact interface but typically rely on exfoliated flakes, transfers, or co-growth [12], which are not scalable to wafer-level integration.

Atomic layer deposition (ALD) offers a scalable alternative, providing conformal coverage, wafer-scale uniformity, and angstrom-level thickness control [13]. ALD can incorporate metal precursors to form degenerately-doped TMD alloys. Nb has proven effective as a substitutional dopant in WSe$_2$ and MoS$_2$ [14], and Nb-enriched WS$_2$ grown by ALD has shown high conductivity with hole densities exceeding 10$^{20}$ cm$^{-3}$ [15],[16]. Such films act as high-work-function, p$^+$-type layers that can transfer charge into adjacent semiconductors, enabling efficient hole injection. Despite these advantages, ALD-based doped interlayers have not yet demonstrated in WSe$_2$ PFETs.

In this work, we demonstrate a scalable, self-aligned contact doping strategy using plasma-enhanced ALD (PEALD) of Nb-doped WS$_2$ (Nb:WS$_2$) as a raised source/drain contact layer. Devices with contact spacing, $L_{DS}$, of 0.15 μm as global active current of > 100 μA/μm at $V_{DS}$ = −1 V and $n_s$ = 7.5 × 10$^{12}$ cm$^{-2}$. A self-aligned contact process using O$_2$ annealing ensures precise alignment of Nb:WS$_2$ to Pd contacts while maintaining excellent channel transport. The use of PEALD as a contact strategy is a promising path toward a manufacturing-compatible 2D transistor process.

## II. DEVICE FABRICATION

The device fabrication started with degenerately doped p$^{++}$ Si wafers with 90-nm thermally grown SiO$_2$, used as global back-gate substrates. Prior to transferring exfoliated WSe$_2$, the substrates were cleaned by sequential rinsing in acetone, isopropanol, and deionized water, dried with N$_2$, and exposed to brief O$_2$ plasma to remove organics. WSe$_2$ flakes of monolayer (1L), bilayer (2L), and few-layer (FL) thicknesses were exfoliated from bulk crystals and transferred onto the substrates. Optical microscopy identified suitable flakes, and thicknesses were verified by Raman spectroscopy. A conformal Nb:WS$_2$ (Nb/W ratio ≈ 0.43) was then deposited by PEALD at 300 °C [15]. Each super-cycle consisted of four W cycles and one Nb cycle, yielding a ~1.6-nm-thick film coating the WSe$_2$ and surrounding oxide.

Device fabrication proceeded by defining transistor mesas with SF$_6$/O$_2$ reactive ion etching and patterning Pd (20 nm)/Au (50 nm) contacts by electron-beam lithography, evaporation and lift-off. The backside oxide was etched in buffered oxide etchant for 2 min to expose the p$^{++}$ Si for to allow backgate contact. FIG. 1 shows a schematic of the fabrication sequence.

Samples were characterized before and after O$_2$ annealing at 200 °C for 20 min. During the annealing step, the Nb:WS$_2$ in the channel regions oxidizes forming an amorphous insulating layer [17], as confirmed by electrical tests in regions without WSe$_2$, while the film beneath the contacts remained intact,

This work was partially funded by Intel Corporation, and also by the University of Minnesota College of Science and Engineering. Portions of this work were conducted in the Minnesota Nano Center, which is supported by the NSF through the National Nanotechnology Coordinated Infrastructure under Award No. ECCS-2025124, and the University of Michigan Lurie Nanofabrication Facility. S. Mishra is with the Dept. of Electrical & Computer Engineering, University of Minnesota, Minneapolis, MN, USA A. A. Adhikari, and S. J. Koester are with the Dept. of Electrical Engineering, University of Notre Dame, Notre Dame, IN, USA (e-mail: skoester@nd.edu). A. A. Bol, L. G. Cooper, and R. A. Dawley are with the Dept. of Chemistry, University of Michigan, Ann Arbor, MI, USA. D. Seo is now with Argonne National Laboratory, R. Li is now with Micron Technology.

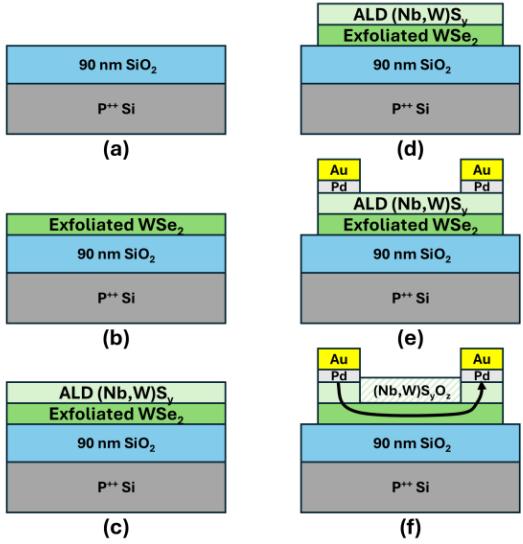

FIG. 1. Process flow for WSe₂ p-MOSFETs. (a) Starting substrate, (b) exfoliation of WSe₂, (c) PEALD deposition of Nb:WS₂, (d) mesa patterning, (e) lift-off of Pd/Au ohmic contact metallization, (f) selective oxidation of Nb:WS₂. The arrow shows the current transport path after the oxidation step.

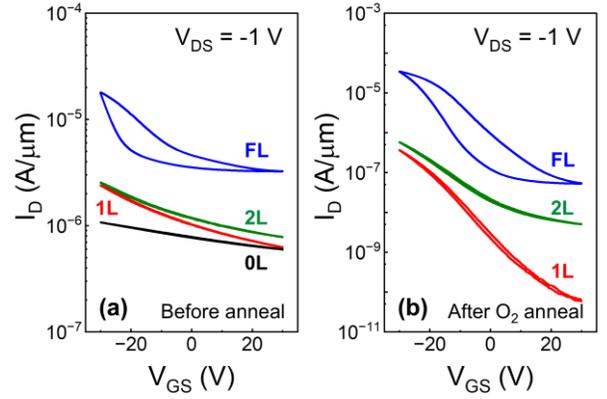

FIG. 2. (a) Plot of drain current, $I_D$, vs. gate-to-source voltage, $V_{GS}$, for 0L, 1L, 2L and FL WSe₂ p-MOSFETs before oxidation of the Nb:WS₂ layer. (b) $I_D$ vs. $V_{GS}$ for same devices in (a) after RTA/O₂ at 200 °C for 20 min. The curves for the 0L devices had current below the noise floor and are not shown. For all devices, $L_{DS} = 0.6$ μm, and data was taken at room temperature.

creating self-aligned doped source/drain regions. Electrical characterization before and after RTA/O₂ was performed at room temperature in a vacuum probe station (<10⁻⁵ Torr) using a Keysight B1500A semiconductor parameter analyzer, with the p⁺⁺ Si substrate as the back gate. Transfer and output characteristics were recorded before and after annealing for direct comparison across 1L, 2L, and FL WSe₂ devices.

## III. RESULTS AND DISCUSSION

FIG. 2 shows the results for devices with $L_{DS} = 0.6$ μm with 1L, 2L and FL WSe₂ channels both before and after O₂ annealing. We also include the results for a device where the Nb:WS₂ was deposited directly on the SiO₂ substrate, and we label this device as 0L. Before anneal (FIG. 2(a)) it is clear that no devices have strong turn-off behavior. This is expected given that the Nb:WS₂ shorts the source and drain together and is only poorly modulated by the bottom-gate electrode. However, we do note that the results are affected by the number of WSe₂ layers, with the on-state current increasing as the number of WSe₂ layers is increased. As we described in previous work, this is strong evidence that charge transfer is taking place between the Nb:WS₂ layer and the WSe₂ channel [16]. Since transport is dominated by the bottom-most WSe₂ layer due to the backgating effect, greater number of layers increases the distance between the channel and the polycrystalline Nb:WS₂ layer. We also note that the FL device has much larger hysteresis than the other devices. While the origin of this is not completely clear, we believe this is due to charge trapping in the WSe₂ itself.

After annealing, the devices show very different behavior as shown in FIG. 2(b). First of all, we find that the 0L samples become completely insulating with current level below the noise floor of our measurements, and so these results are not shown. For 1L-WSe₂, the on-off current ratio increases substantially from ~3 to nearly $6 \times 10^3$. This is attributed to the oxidation of the Nb:WS₂ between the S/D, which renders it insulating, thus removing the shunt leakage path between the S/D contacts. However, the on-state current is also reduced dramatically, decreasing by roughly a factor of 10. A similar situation is observed for the 2L-WSe₂ sample, though the on-off current ratio is lower than for 1L-WSe₂. Since the Nb:WS₂ is deposited using PEALD, we attribute the reduced performance to mobility degradation of the channel due to plasma-induced damage.

After oxidation, the result for the FL-WSe₂ sample is very different. Here, the on-state current actually increases after O₂ annealing, going from 18 μA/μm to 34 μA/μm. Due to much larger thickness of the WSe₂ channel, we believe the bottom WSe₂ is protected from the plasma damage, allowing the transport properties to be closer to the values for pristine material. The on-off current ratio for the FL-WSe₂ device is also improved, going from ~ 5.5 to over 600.

Given the improved performance of the FL-WSe₂ devices, we performed additional experiments to study the contact-spacing dependence and the results are shown in FIG. 3. Here, devices with $L_{DS} = 0.6, 0.4, 0.2$ and $0.15$ μm were tested. Before O₂ annealing, the devices showed similar behavior to those in FIG. 2. After annealing, the devices show increasing drive current, with values of 34, 43, 96, and 103 μA/μm, at $V_{GS} = -30$ V and $V_{DS} = -1$ V (FIG. 3(a)). The devices all show similar hysteresis and turn-off behavior, though the on-off current ratio decreases slightly with shorter channel length (FIG. 3(b)). The transconductance, $g_m$, also increases going from $L_{DS}$ = 0.6 to 0.2 μm, but then decreases slightly at $L_{DS}$ = 0.15 μm (FIG. 3(c)). This could be a result of short-channel effects, but may also indicate the onset of contact-limited performance.

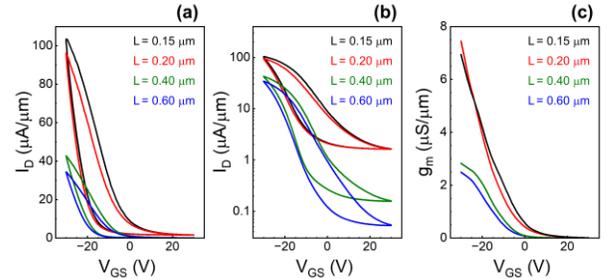

FIG. 3. (a) Plot of drain current, $I_D$, vs. gate-to-source voltage, $V_{GS}$, on a linear scale for FL WSe₂ p-MOSFETs with different source/drain contact spacing, $L_{DS}$, at $V_{DS} = -1$ V. (b) Same plot as in (a) except on a semi-log plot. (c) Transconductance, $g_m$, vs. $V_{GS}$, where $g_m$ is averaged between up and down sweeps.

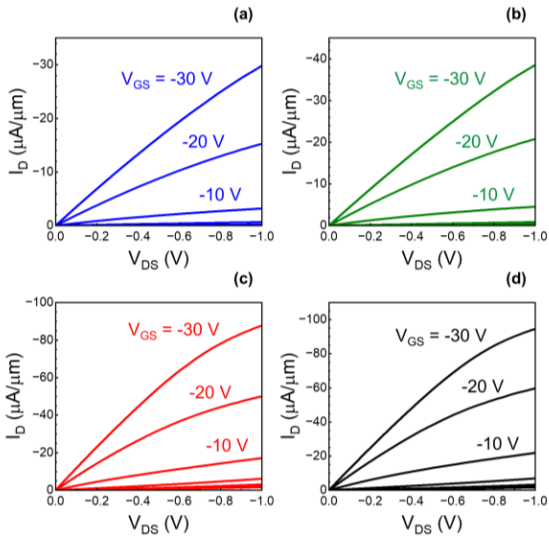

FIG. 4. Output characteristics for FL WSe$_2$ p-MOSFETs with source/drain contact spacing, $L_{DS}$, of (a) 0.6 μm, (b) 0.4 μm, (c) 0.2 μm, and (d) 0.15 μm.

The output characteristics of the same devices are shown in FIG. 4. All devices show linear turn-on, indicative of ohmic contacts. The on-state currents are slightly different than those in FIG. 3, which we attribute to the hysteretic nature of the backgate, which often can cause variations in current depending upon how the device is swept. For the $L_{DS}$ = 0.15-μm device, the average threshold voltage between up and down sweeps is +1 V, and so at $V_{GS}$ = −30 V, $n_s$ = 7.5 × 10$^{12}$ cm$^{-2}$. Furthermore, based upon the maximum $g_m$ value in FIG. 3(c), we extract a field-effect mobility of 21 cm$^2$/Vs for the $L_{DS}$ = 0.6 μm device.

Our results compare favorably with previous reports in the literature on WSe$_2$ p-MOSFETs [18]-[37], as summarized in FIG. 5. Here, we have plotted our figure of merit (FOM), $I_{ON}$ / $n_s$ vs. gate length, where our FOM accounts for devices under different gating conditions. Our results compare favorably with most reports in the literature, providing strong evidence of the benefit of ALD-based doping for contact formation, particularly given the CMOS-compatibility of our ALD scheme relative to other techniques.

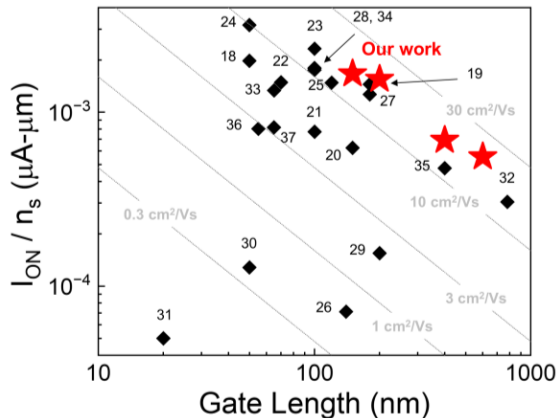

FIG. 5. Plot at $V_{DS}$ = −1 V showing our work (red stars) compared to WSe$_2$ p-MOSFETs in the literature [18]-[37]. Gray dashed lines show expected trends for $I_{ON}/n_s$ at different mobility values.

## IV. CONCLUSION

In conclusion, we have demonstrated a technique to form contacts to 2D p-MOSFETs using Nb:WS$_2$ contacts formed using PEALD. The devices are fabricated using a technique that aligns the contact metallization with the Nb:WS$_2$ using a selective oxidation process. Devices with source/drain spacing of 0.15 μm have on-state current of 103 μA/μm at $V_{DS}$ = −1 V and $n_s$ ~ 7.5 × 10$^{12}$ cm$^{-2}$. The work demonstrates the potential of ALD TMDs as a contact solution for ultra-scaled CMOS.